\documentclass[aps,prl,numerical,linenumbers,superscriptaddress,showpacs,floatfix,reprint]{revtex4-1}
\usepackage[dvipdfm]{graphicx}
\usepackage{amsmath,amssymb,amsfonts}
\usepackage{color}
\usepackage[colorlinks=true,linkcolor=blue,plainpages=false]{hyperref}

\def\v2{\mbox{$v_2$}}

\newcommand{\mean}[1]{\left\langle #1 \right\rangle}

\begin{document}
%


\title{ Initial indications for the production of a strongly coupled plasma \\
in Pb+Pb collisions at $\sqrt{s_{NN}} = 2.76$ TeV
}
%
%
\author{ Roy~A.~Lacey}
\email[E-mail: ]{Roy.Lacey@Stonybrook.edu}
\affiliation{Department of Chemistry, 
Stony Brook University, \\
Stony Brook, NY, 11794-3400, USA}
\author{A. Taranenko}
\affiliation{Department of Chemistry, 
Stony Brook University, \\
Stony Brook, NY, 11794-3400, USA}
\author{ N. N. Ajitanand} 
\affiliation{Department of Chemistry, 
Stony Brook University, \\
Stony Brook, NY, 11794-3400, USA}
\author{ J. M. Alexander}
\affiliation{Department of Chemistry, 
Stony Brook University, \\
Stony Brook, NY, 11794-3400, USA}
%
%
%
%

\date{\today}


\begin{abstract}
 
	Results from first measurements of charged particle differential elliptic flow, 
obtained in Pb+Pb collisions at $\sqrt{s_{NN}} = 2.76$ TeV with the ALICE detector 
at  CERN's Large Hadron Collider (LHC), are compared to those obtained for Au+Au 
collisions at $\sqrt{s_{NN}} = 0.2$ TeV with the PHENIX detector at BNL's 
Relativistic Heavy Ion Collider (RHIC). The comparisons, made as a function 
of centrality (cent) or the number of participant pairs ($N_{\text{part}}$) 
and particle transverse momentum $p_T$, indicate an excellent agreement between 
the magnitude and trends for the flow coefficients ${v_2(p_T,\text{cent})}$. 
Analysis indicates that the averaged specific viscosity of the quark gluon 
plasma (QGP) produced in LHC collisions, is similar to that for the strongly 
coupled QGP produced in RHIC collisions. 
\end{abstract}
\maketitle

	
	First results from Pb+Pb collisions at $\sqrt{s_{NN}} = 2.76$ TeV,
from CERN's Large Hadron Collider (LHC) \cite{Aamodt:2010pb,Aamodt:2010pa}
have initiated the highly anticipated explorations of the 
the high temperature, high entropy density domain of the QCD phase diagram.
At $\sim 14$ times the energy of RHIC collisions, these Pb+Pb collisions 
are expected to create a rapidly thermalized plasma of quarks and gluons 
(QGP) at temperatures higher than those currently accessible at RHIC. 
The reported hadron multiplicity in these Pb+Pb collisions is 
$\sim 1584$ (or 8.3 per participating nucleon pair $N_{\text{part}}$) for 
the most central 5\% of the hadronic cross 
section \cite{Aamodt:2010pb} -- a factor of 2.2 increase over that observed 
in central Au+Au collisions at RHIC ($\sqrt{s_{NN}} = 0.2$ TeV). Thus, it 
appears that one now has a lever arm for probing the QGP's viscosity and other 
transport properties to determine if they evolve from the strongly coupled plasma 
observed at RHIC \cite{Gyulassy:2004vg,Gyulassy:2004zy,Lacey:2005qq,Shuryak:2008eq,Lacey:2009xx}, 
towards the more weakly interacting, gaseous plasma state expected at asymptotically 
high temperatures.

	In non-central heavy ion collisions, the spacial asymmetry of an initial ``almond-shaped" 
collision-zone leads to {\it flow}. That is, partonic interactions in this collision-zone drive 
uneven pressure gradients in- and out of the reaction plane and hence, a momentum anisotropy 
of the particles emitted about this plane.
At mid-rapidity, the magnitude of this flow is frequently characterized with the even-order 
Fourier coefficients;
%
%
$
v_{\rm n} = \mean{e^{in(\Delta\phi)}}, {\text{  }} n=2,4,..., 
$
%
%
where $\Delta\phi$ is the azimuth of an emitted hadron 
about the reaction plane, and brackets denote 
averaging over particles and events.

	Because they are known to be sensitive to various transport properties of the 
expanding hot medium \cite{Heinz:2002rs,Teaney:2003kp,Lacey:2006pn,
Romatschke:2007mq,Song:2007ux,Drescher:2007cd,Xu:2007jv,Greco:2008fs,
Luzum:2008cw,Chaudhuri:2009hj}, the differential Fourier coefficients $v_{2}(N_{\text part})$,  
$v_{2}({p_T})$ and $v_{2}(N_{\text part}, p_T)$ have been extensively studied 
as a function of collision centrality (cent) and hadron transverse momentum $p_T$,
in Au+Au collisions at 
RHIC ($\sqrt{s_{NN}} = 0.06 - 0.2$ TeV) \cite{Adcox:2002ms,Adams:2003zg,Adler:2003kt,Adler:2004cj,Alver:2006wh,
Afanasiev:2007tv,Star:2008ed,Afanasiev:2009wq,Adare:2010ux}. 
Indeed, considerable effort is currently being devoted to the quantitative extraction 
of the specific shear viscosity $\eta/s$ ({\em i.e.}\ the ratio 
of shear viscosity $\eta$ to entropy density $s$) via comparisons to viscous relativistic 
hydrodynamic simulations \cite{Luzum:2008cw,Song:2008hj,Dusling:2007gi,
Chaudhuri:2009hj,Bozek:2009mz,Peschanski:2009tg,Denicol:2010tr,Holopainen:2010gz,Schenke:2010rr,
Song:2010mg}, 
transport model calculations \cite{Molnar:2001ux,Xu:2007jv,Greco:2008fs} and hybrid approaches which 
involve the parametrization of scaling deviations from ideal hydrodynamic behavior \cite{Lacey:2006pn,Drescher:2007cd,Lacey:2009xx,Masui:2009pw,Lacey:2010fe}. 

With the advent of detailed $v_{2}(\text{cent}, p_T)$ data for 
Pb+Pb collisions at the LHC ($\sqrt{s_{NN}} = 2.76$ TeV), an important 
question is  whether these new flow data give an early indication for a significant 
difference in the viscosity of the QGP produced in RHIC and LHC collisions?
Such a difference might be expected because, relative to Au+Au collisions at RHIC, 
the measured multiplicity for Pb+Pb collisions at $\sqrt{s_{NN}} = 2.76$ TeV, suggests an 
approximate 30\% increase in the temperature of the QGP produced in LHC collisions. 
	
\begin{figure}[t]
\includegraphics[width=1.0\linewidth]{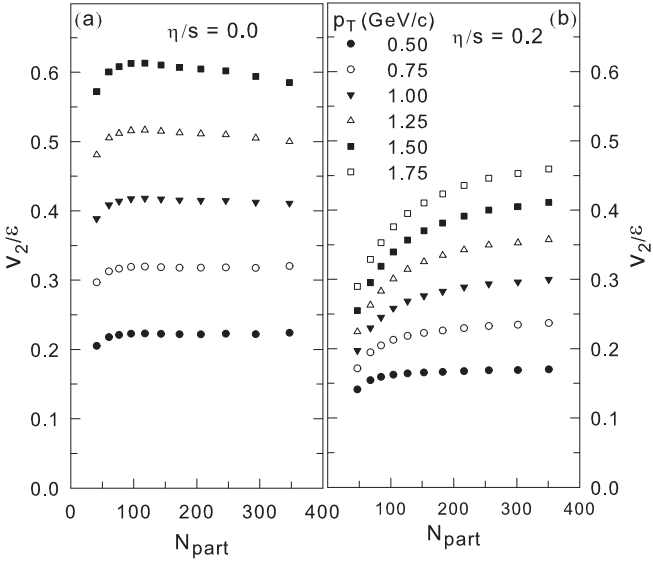} 
\caption{(color online) Comparison of $v_2/\varepsilon_2$ vs. $N_{\rm part}$ 
for several $p_T$ selections, obtained from perfect fluid (a) and 
viscous (b) hydrodynamic simulations of Au+Au collisions. For these calculation, 
a Glauber initial eccentricities are use in conjunction with a lattice-based 
equation of state \cite{Dusling:2009df}.  
}
\label{Fig1}
\end{figure}

	The influence of $\frac{\eta}{s}$ on anisotropic flow is especially transparent 
in studies involving the flow coefficient scaled by the initial eccentricity of 
the collision zone $\frac{v_{2}(N_{\text{part}}, p_T)}{\varepsilon_{2}(N_{\text{part}})}$, 
as illustrated in Fig. \ref{Fig1}. Here, results from hydrodynamic 
simulations (with the code of Dusling and Teaney \cite{Dusling:2009df}) are shown 
for two different viscosity values. For $\frac{\eta}{s} = 0$, 
Fig. \ref{Fig1} (a) indicates an essentially flat dependence for 
$\frac{v_{2}(N_{\text{part}}, p_T)}{\varepsilon_{2}(N_{\text{part}})}$ in 
line with the expected scale invariance of perfect fluid hydrodynamics.
By contrast, Fig. \ref{Fig1} (b) shows that the introduction of a viscosity 
($\frac{\eta}{s} = 0.2$) reduces the magnitude of $v_{2}(N_{\text{part}}, p_T)$ 
and breaks the scale invariance of ideal hydrodynamics evidenced 
in Fig. \ref{Fig1} (a). That is, there are substantial $p_T$-dependent deviations 
away from the essentially flat $N_{\text{part}}$ dependence observed 
in Fig. \ref{Fig1} (a).
%

\begin{figure}[t]
\includegraphics[width=1.0\linewidth]{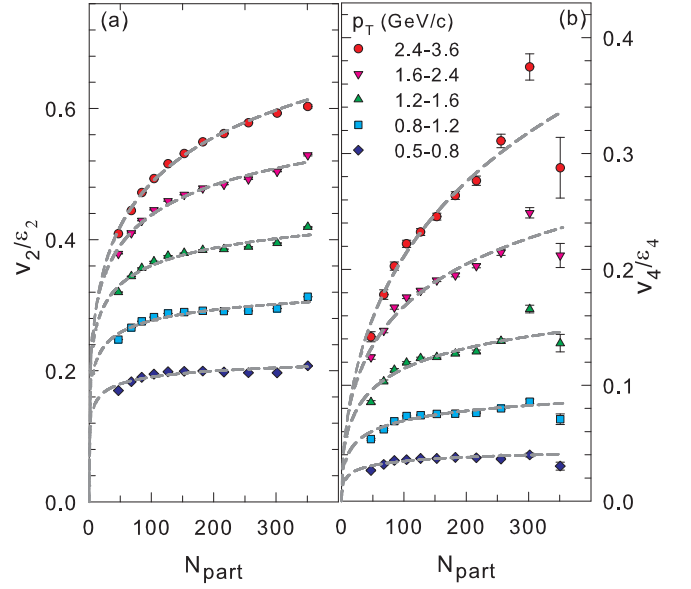} 
\caption{(color online) Comparison of $v_2/\varepsilon_2$ vs. $N_{\rm part}$ (a) 
and $v_4/\varepsilon_4$ vs. $N_{\rm part}$ (b) for several $p_T$ selections as 
indicated. The dashed curves indicate a simultaneous fit to 
the data in (a) and (b) [for each $p_T$] \cite{Lacey:2010fe}. 
The $v_{2,4}$ data are from Ref. \cite{Adare:2010ux}. 
}
\label{Fig2}
\end{figure}

	Figure \ref{Fig2} shows that these predicted scaling deviations are found 
in actual experimental data \cite{Lacey:2010fe}. It shows eccentricity-scaled values of 
$v_{2,4}(p_T,N_{\text{part}})$ (obtained with factorized Kharzeev-Levin-Nardi [MC-KLN] 
model eccentricities \cite{Lappi:2006xc,Drescher:2007ax}) for several $p_T$ cuts. 
The low-$p_T$ selections show small scaling deviations, {\it i.e.} they are almost flat.
However, the data points slope upward progressively (from low to high $N_{\rm part}$) 
as the $\left< p_T \right>$ is increased, reflecting an increase in the scaling 
deviations with $\left< p_T \right>$. 

	These eccentricity-scaling deviations reflect the effects of viscosity, as well 
as its attendant influence on the emission distribution ($f$) on the freeze-out surface. 
This distribution can be expressed as \cite{Teaney:2003kp,Dusling:2009df};
\begin{equation}
\frac{dN}{dy p_T dp_T d\phi} \sim f_0 + \delta f \equiv f_0\left( 1+ C\left(\frac{p_{T}}{T_{\!f}}\right)^{2-\alpha}\right),
%
\label{eq2}
\end{equation}  
where $f_0$ is the equilibrium distribution, $T_{\!f}$  is the freeze-out temperature, 
$C \approx \frac{\eta}{3\tau s T_{\!f}}$ and  $\alpha$ is estimated to be 
0 \cite{Lacey:2010fe}; $\tau$ is the time scale of the expansion.
Note that the factor $\delta f$ results [explicitly] from a finite shear viscosity and 
is known to dominate the calculated viscous corrections to $v_2(p_T)$ for 
$p_T \agt 1$ GeV/c due to its strong $p_T^2$ dependence \cite{Dusling:2009df}.
Thus, a significant increase in the value of $\frac{\eta}{s}$ would not only serve 
to decrease the magnitude of $\frac{v_{2}(N_{\text{part}}, p_T)}{\varepsilon_{2}(N_{\text{part}})}$
but would also magnify the eccentricity-scaling deviations, especially for $p_T \agt 1$ GeV/c. 
%

\begin{figure}[t]
\includegraphics[width=0.75\linewidth]{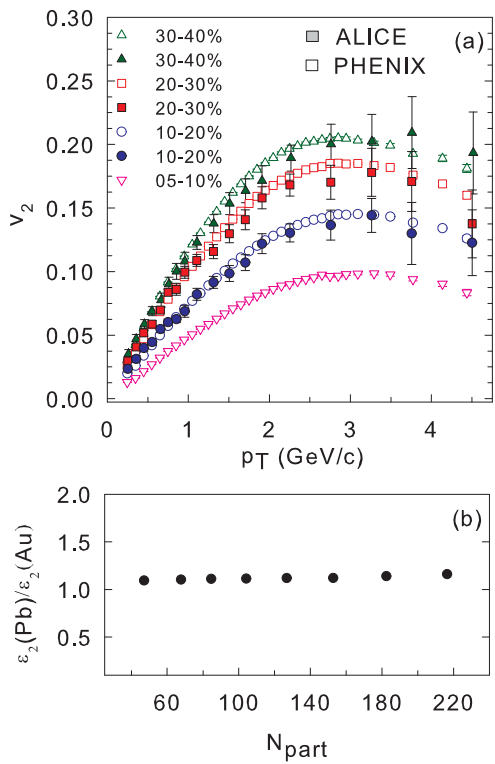}
\caption{(color online) Comparison of $v_2$ vs. $p_T$ 
for several centrality selections as indicated (a). The ALICE and PHENIX 
data are from Refs. \cite{Aamodt:2010pa} and \cite{Adare:2010ux} respectively. 
The ratio of the initial eccentricity for Pb+Pb and Au+Au collisions is 
shown as a function of $N_{\rm part}$ in panel (b).
}
\label{Fig3}
\end{figure}

	Figures \ref{Fig1} and \ref{Fig2} show that a simple way to test for a change 
in $\frac{\eta}{s}$ for two different data sets, is to compare their respective 
eccentricity-scaled anisotropy coefficients   
$\frac{v_{2}(N_{\text{part}}, p_T)}{\varepsilon_{2}(N_{\text{part}})}$ 
and $\frac{v_{4}(N_{\text{part}}, p_T)}{\varepsilon_{4}(N_{\text{part}})}$, to see if 
they differ. That is, a significant $\frac{\eta}{s}$ difference would not only lead to 
different magnitudes, but also to very different $p_T$-dependent curvatures for the 
eccentricity-scaled coefficients from each data set. 
If the $N_{\text{part}}$ dependence of ${\varepsilon_{2,4}}$ is the same 
for both data sets, then the test can be made more simple by directly 
comparing the flow coefficients ${v_{2}({\text{cent}}, p_T)}$.
Indeed, the calculated MC-KLN initial eccentricities for the two reactions are very similar 
as shown in Fig. \ref{Fig3} (b). The same trend is observed for Glauber initial eccentricities which are smaller than the MC-KLN values. The ratios in Fig. \ref{Fig3} (b) are a little larger than unity due 
to the larger size of the Pb nucleus. However, for the same 
centrality, they are $\approx 1$ as also noted in Ref. \cite{Aamodt:2010pa}.

	The flow results recently reported in Ref. \cite{Aamodt:2010pa} have also indicated a 
strong similarity between the elliptic flow coefficients $v_2({\text{cent}}, p_T)$ obtained 
by the ALICE collaboration for Pb+Pb collisions at $\sqrt{s_{NN}} = 2.76$ TeV and 
those obtained by the STAR collaboration for Au+Au collisions at $\sqrt{s_{NN}} = 0.2$ TeV.
Given that the differences between the Glauber-based initial eccentricities  
for Au+Au and Pb+Pb collisions are small for the same centrality selection  
(cf. Fig. \ref{Fig3} and Ref. \cite{Aamodt:2010pa}), the measured 
flow coefficients for both data sets can be directly compared to test 
for a viscosity difference. 

\begin{figure}[t]
\includegraphics[width=0.8\linewidth]{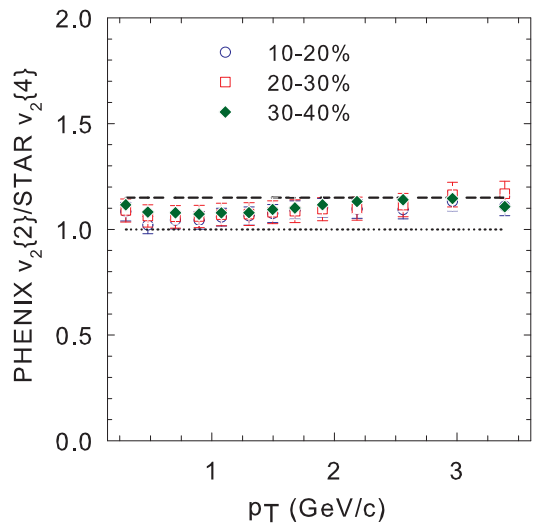}
\caption{(color online) Comparison of PHENIX's $v_2\{2\}$ vs. $p_T$ 
and STAR's $v_2\{4\}$ vs. $p_T$
for several centrality selections as indicated. The STAR and PHENIX 
data are from Refs. \cite{Aamodt:2010pa} and \cite{Adare:2010ux} respectively. 
The dotted and dashed lines indicate ratios of 1.0 and 1.15 respectively.
}
\label{Fig4}
\end{figure}

	A comparison of $v_{2}(p_T)$ for several centrality selections from 
the PHENIX  \cite{Adare:2010ux} and ALICE  \cite{Aamodt:2010pa} data sets, is 
shown in Fig. \ref{Fig3} (a). 
The comparison shows good agreement between the magnitudes and trends for both data sets, indicating a strong similarity between the viscous corrections to $v_2(p_T)$ in Pb+Pb ($\sqrt{s_{NN}} = 2.76$ TeV) and Au+Au ($\sqrt{s_{NN}} = 0.2$ TeV) collisions \cite{recent_hydro}. Parenthetically, an exact agreement between the magnitudes of both data sets 
is not to be expected because the ALICE measurements were obtained via the 4-particle cumulant method \cite{Borghini:2001vi} while the PHENIX 
measurements were obtained via the event plane method, albeit with a sizable 
$\Delta\eta$-separation between the event plane and the detected hadrons \cite{Adare:2010ux}. These different measuring techniques reflect different 
associated eccentricity fluctuations which manifest as a small difference 
in the magnitudes of the two data sets. This difference is illustrated in Fig. \ref{Fig4} where we show the ratio of the PHENIX $v_2\{2\}$ measurements to STAR's four particle $v_2\{4\}$ measurements. The ratios show the expected 9-12\% difference(esentially independent of $p_T$) due to the larger inherent fluctations for the $v_2\{2\}$ measurements \cite{Bhalerao:2006tp,Taranenko:2011zz}. This difference does not alter the arguments nor the conclusions which follow. 
The observed agreement between the $v_{2}(p_T)$ data from both the LHC and RHIC implies that the observed increase of the $p_T$-integrated $v_2$ (from RHIC to the LHC) \cite{Aamodt:2010pa}, can be 
simply explained by an increase in the $\left\langle p_T \right\rangle$. 
%

As in Refs. \cite{Lacey:2009xx,Lacey:2010fe} the deviations from eccentricity-scaling have  
been used to characterize the magnitude of the viscous corrections to  
$\frac{v_{2}(N_{\text{part}}, p_T)}{\varepsilon_{2}(N_{\text{part}})}$ and 
$\frac{v_{2}(N_{\text{part}})}{\varepsilon_{2}(N_{\text{part}})}$
\cite{Bhalerao:2005mm,Lacey:2006pn,Drescher:2007cd,Masui:2009pw}
by a Knudsen number ($K ={\lambda}/{\bar R}$) 
parametrization, where ${\lambda}$ is the mean free path and ${\bar R}$
is the transverse size of the system obtained from the same Glauber-based calculations 
used to determine ${\varepsilon_{2}(N_{\text{part}})}$.
In turn, the extracted Knudsen number 
provides an estimate for the specific viscosity of the QGP;
\begin{equation}
\frac{\eta}{s} \approx \lambda T c_s \equiv (\bar{R}KT c_s),
\label{eq:2}
\end{equation}
where $c_s$ is the sound speed estimated from lattice 
calculations \cite{Huovinen:2009yb} for the mean temperature $T$. 
The agreement between the LHC and RHIC data shown in Fig. \ref{Fig3} (a) and 
in Fig. 2 of Ref. \cite{Aamodt:2010pa}, indicate very similar viscous corrections and 
thus, a similar $\frac{\eta}{s}$ range for the plasma produced at higher 
temperatures in Pb+Pb collisions at $\sqrt{s_{NN}} = 2.76$ TeV.
In Ref. \cite{Lacey:2010fe} the estimate $4\pi\frac{\eta}{s} \sim 1 - 2$ was 
obtained for the K values extracted using MC-KLN and MC-Glauber eccentricities 
[respectively] in central and mid-central Au+Au collisions ($\sqrt{s_{NN}} = 0.2$ TeV) 
for the mean temperature $T=220 \pm 20$ MeV \cite{Adare:2008fqa}. 

	The similarity between the $\frac{\eta}{s}$ values for 
the plasma produced in RHIC and LHC collisions can be understood in the 
framework of Eq. \ref{eq:2}, via the following simple estimate for the 
Knudsen number \cite{Danielewicz:1984ww,Arnold:2000dr};
\begin{equation}
   K = \left( \frac{\beta}{\bar{R}T}\right),
\end{equation}
where the magnitude of $\beta$ depends primarily on whether the plasma is 
strongly or weakly coupled (for a weakly couple plasma, $\beta \sim 36/8.144g^4$).  
Substitution of the estimate for $K$ into Eq. \ref{eq:2}
shows that very little change in $\frac{\eta}{s}$ would result if  
the coupling strength of the plasma remains essentially the same for two 
different mean temperatures, {\em i.e.} the mean sound speed does not 
show a strong temperature dependence over the range of interest.
Note that a similar argument applies for the comparison of RHIC differential 
$v_2$ data over the beam collision energy range 
$\sqrt{s_{NN}} = 0.062 - 0.2$ TeV, where $v_2(p_T, \text{cent})$ has been 
observed to be approximately constant for Au+Au collisions \cite{Adler:2004cj}. 
Here, an important difference is that the associated temperature change 
is relatively small.


In summary, we have made detailed comparisons between measurements of charged 
particle differential elliptic flow obtained in Pb+Pb collisions at 
$\sqrt{s_{NN}} = 2.76$ TeV, and those obtained for Au+Au 
collisions at $\sqrt{s_{NN}} = 0.2$ TeV with the PHENIX detector at RHIC. 
The comparisons indicate an excellent agreement between the magnitude and 
trends for the flow coefficients ${v_2(p_T,{\text{cent}})}$. Our analysis 
indicates that the averaged specific viscosity of the QGP produced in LHC 
collisions is similar to that for the strongly coupled QGP produced in RHIC collisions. 
Therefore, a strong indication for an evolution toward a more weakly interacting 
plasma has not been exhibited.
It will be most interesting to investigate whether or not this conclusion is 
further supported by detailed viscous hydrodynamic calculations, as well as 
more detailed differential flow measurements at the LHC.

{\bf Acknowledgments:}
We thank R. Snellings for providing the ALICE experimental data.
This research is supported by the US DOE under contract DE-FG02-87ER40331.A008.
 


%
\bibliography{Lacey_LHC_viscosity} 
\end{document}